\documentclass[lettersize,journal]{IEEEtran}
\usepackage{amsmath,amsfonts}
\usepackage{algorithm}
\usepackage{array}
\usepackage[caption=false,font=normalsize,labelfont=sf,textfont=sf]{subfig}
\usepackage{textcomp}
\usepackage{stfloats}
\usepackage{url}
\usepackage{verbatim}
\usepackage{graphicx}
\usepackage{cite}
\usepackage{algpseudocode}
\usepackage{amsmath}
\usepackage{listings}
\usepackage{tabularx}
\usepackage{booktabs}
\usepackage{multirow}
\usepackage{graphicx}
\usepackage{bm}
\usepackage{amsmath, amssymb, amsfonts}
\begin{document}

\title{An Ultra-low Power TinyML System for Real-time Visual Processing at Edge}

\author{Kunran Xu\textsuperscript{\dag },
        Huawei Zhang\textsuperscript{\dag},
        Yishi Li,
        Yuhao Zhang,
        Rui Lai,~\IEEEmembership{Member,~IEEE},
        and Yi Liu%

\thanks{\dag\,  Authors contributed equally to this work.}
\thanks{This work was supported in part by the National Key R\&D Program of China under Grant 2018YF70202800, Natural Science Foundation of China (NSFC) under Grant 61674120. (Corresponding author: Rui Lai).}
\thanks{Kunran Xu, Huawei Zhang, Yishi Li, Yuhao Zhang and Rui Lai are with the School of Microelectronics, Xidian University, Xi’an 710071, and also with the Chongqing Innovation Research Institute of Integrated Cirtuits, Xidian University, Chongqing 400031, China. (e-mail: aazzttcc@gmail.com; myyzhww@gmail.com; yshlee1994@outlook.com; stuyuh@163.com; rlai@mail.xidian.edu.cn).}
\thanks{Yi Liu is with the School of Microelectronics, Xidian University, Xi'an 710071, and also with the Guangzhou Institute of Technology, Xidian University, Guangzhou 510555, China. (yiliu@mail.xidian.edu.cn).}}

\markboth{Journal of \LaTeX\ Class Files,~Vol.~14, No.~8, August~2021}%
{Shell \MakeLowercase{\textit{et al.}}: A Sample Article Using IEEEtran.cls for IEEE Journals}

\maketitle

\begin{abstract}
Tiny machine learning (TinyML), executing AI workloads on resource and power strictly restricted systems, is an important and challenging topic. This brief firstly presents an extremely tiny backbone to construct high efficiency CNN models for various visual tasks. Then, a specially designed neural co-processor (NCP) is interconnected with MCU to build an ultra-low power TinyML system, which stores all features and weights on chip and completely removes both of latency and power consumption in off-chip memory access. Moreover, an application specific instruction-set is further presented for realizing agile development and rapid deployment. Extensive experiments demonstrate that the proposed TinyML system based on our tiny model, NCP and instruction set yields considerable accuracy and achieves a record ultra-low power of 160mW while implementing object detection and recognition at 30FPS. The demo video is available on \url{https://www.youtube.com/watch?v=mIZPxtJ-9EY}.
\end{abstract}


\vspace{-0.2cm}
\section{Introduction}

\IEEEPARstart{R}unning machine learning inference on the resource and power limited environments, also known as Tiny Machine Learning (TinyML), has grown rapidly in recent years. It is promising to drastically expand the application domain of healthcare, surveillance, and IoT, \emph{etc}~\cite{mcunet,tinyml}. However, TinyML presents severe challenges due to large computational load and memory demand of AI models, especially in vision applications. Popular solutions using CPU+GPU architecture has shown high flexibility in MobileML applications~\cite{mittal2019survey}, but it is no longer feasible in TinyML for the much stricter constraints on hardware resources and power consumption. A typical TinyML system based on microcontroller unit (MCU) usually has only $<$ 512KB on-chip SRAM, $<$2MB Flash, $<$1GOP/s computing capability, and $<$1W power limitation~\cite{tinyml, cmsis}. Meanwhile, it is difficult to use off-chip memory (\emph{e.g.}, DRAM) in TinyML system for the very limited energy budget, showing a huge gap between the desired and available storage capacity for running visual AI models.

Recently, the continuously emerging studies on TinyML achieve to deploy CNNs on MCUs by introducing memory-efficient inference engines~\cite{mcunet,cmsis} and more compact CNN models~\cite{micronets,RT1060}. However, the existing TinyML systems still struggle to implement high-accuracy and real-time inference with ultra-low power consumption. Such as the state-of-the-art MCUNet~\cite{mcunet} obtains 5FPS on STM32F746 but only achieves 49.9\% top-1 accuracy on ImageNet. When the frame rate is increased to 10FPS, the accuracy of MCUNet further drops to 40.5\%. What's more, running CNNs on MCUs is still not a extremely power-efficient solution due to the low efficiency of general purpose CPU in intensive convolution computing and massive weight data transmission.
\begin{figure}[t]
\centering
\includegraphics[width=0.9 \linewidth]{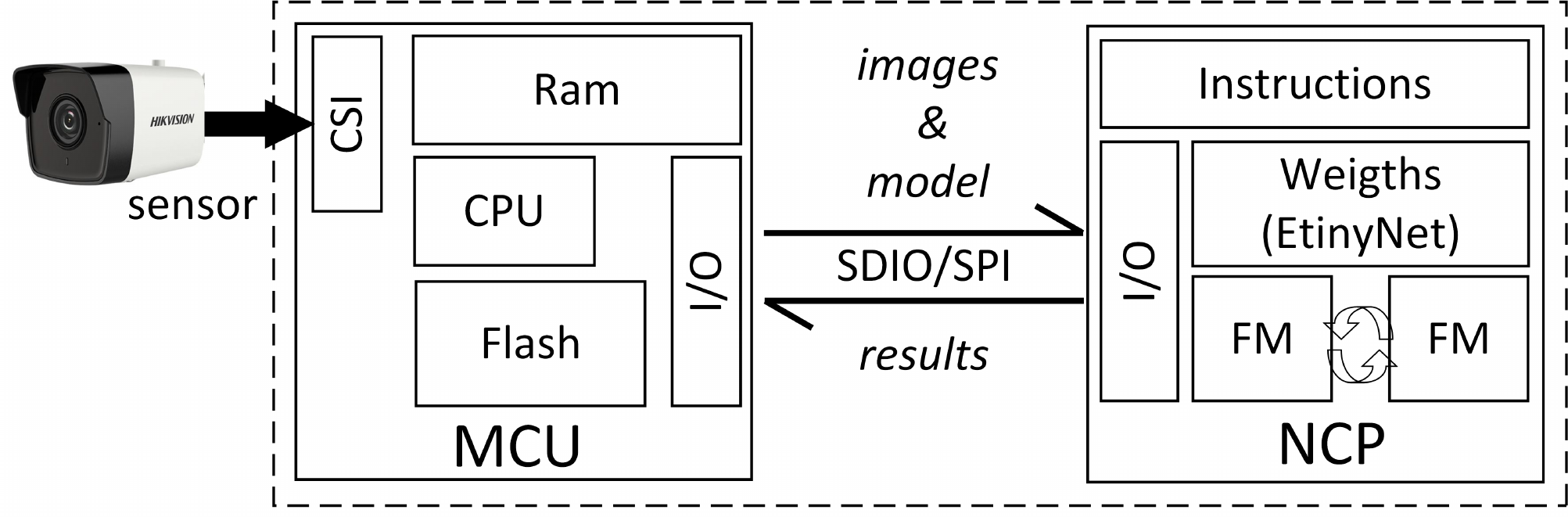}
\vspace{-0.2cm}
\caption{The overview of the proposed TinyML system for visual processing.}
\vspace{-0.2cm}
\label{fig:system}
\vspace{-0.2cm}
\end{figure}
Considering this, we propose to greatly pormote TinyML system by jointly designing more efficient CNN models and specific CNN co-processor. Specifically, we firstly design an extemelly tiny CNN backbone EtinyNet aiming at TinyML applications, which has only 477KB model weights and  maximum feature map size of 128KB and still yields remarkable 66.5\% ImageNet Top-1 accuracy. Then, an ASIC-based neural co-processor (NCP) is specially designed for accelerating the inference. Since implementing CNN inference in a fully on-chip memory access manner, the proposed NCP achieves up to 180FPS throughput with 73.6mW ultra-low power consumption. On this basis, we propose a state-of-the-art TinyML system shown in Fig.\ref{fig:board} for visual processing, which yields a record low power of 160mW in object detecting and recognizing at 30FPS.

In summary, we make the following contributions:
\begin{itemize}
\item[1)] An extremely tiny CNN backbone named EtinyNet is specially designed for TinyML. It is far more efficient than existing lightweight CNN models.
\item[2)] An efficient neural co-processor (NCP) with specific designs for tiny CNNs is proposed. While running EtinyNet, NCP provides remarkable processing efficiency and convenient interface with extensive MCUs via SDIO/SPI.
\item[3)] Building upon the proposed EtinyNet and NCP, we promote the visual processing TinyML system to achieve a record ultra-low power and real-time processing efficiency, greatly advancing the TinyML community.
\end{itemize}
\section{Solution of Our TinyML System} Fig.\ref{fig:system} shows the overview of the proposed TinyML system. It integrates MCU with the specially designed energy-efficient NCP on a compact board to achieve superior efficiency in a collaborative work manner. To the best of our knowledge, we are the first to propose such a collaborative architecture in TinyML field, which successfully balances the efficiency and flexibility in the inference.

Initially, MCU sends the model weights and instructions to NCP who has sufficient on-chip SRAM to cache all these data. During inference, NCP computes the intensive CNN backbone efficiently while MCU only performs the light-load pre-processing (color normalization) and post-processing (fully-connected layer, non-maximum suppression, \emph{etc}), which improves the overall energy efficiency to the greatest extent. Besides, the inference process of NCP only involves two kinds of data transfer, which are the input image and the output results. This working mode greatly reduces the off-chip data transfer power consumption and overall processing latency, and helps the system to achieve high energy efficiency in computing, which will be demonstrated in Section VI.

Considering real-time application, we interconnects NCP and MCU with SDIO/SPI interface. SDIO could provide up to 500Mbps bandwidth, which can transmit about 300FPS for $256\times256$ RGB image and 1200FPS for $128\times128$ one. As for SPI, it still reaches 100Mbps, or an equivalent throughput of 60FPS for $256\times256$ RGB image. These two buses are widely supported by MCUs available in the market, which makes NCP can be applied in a wide range of TinyML systems.

\begin{figure}[t]
	\begin{center}
		\includegraphics[width=0.85 \linewidth]{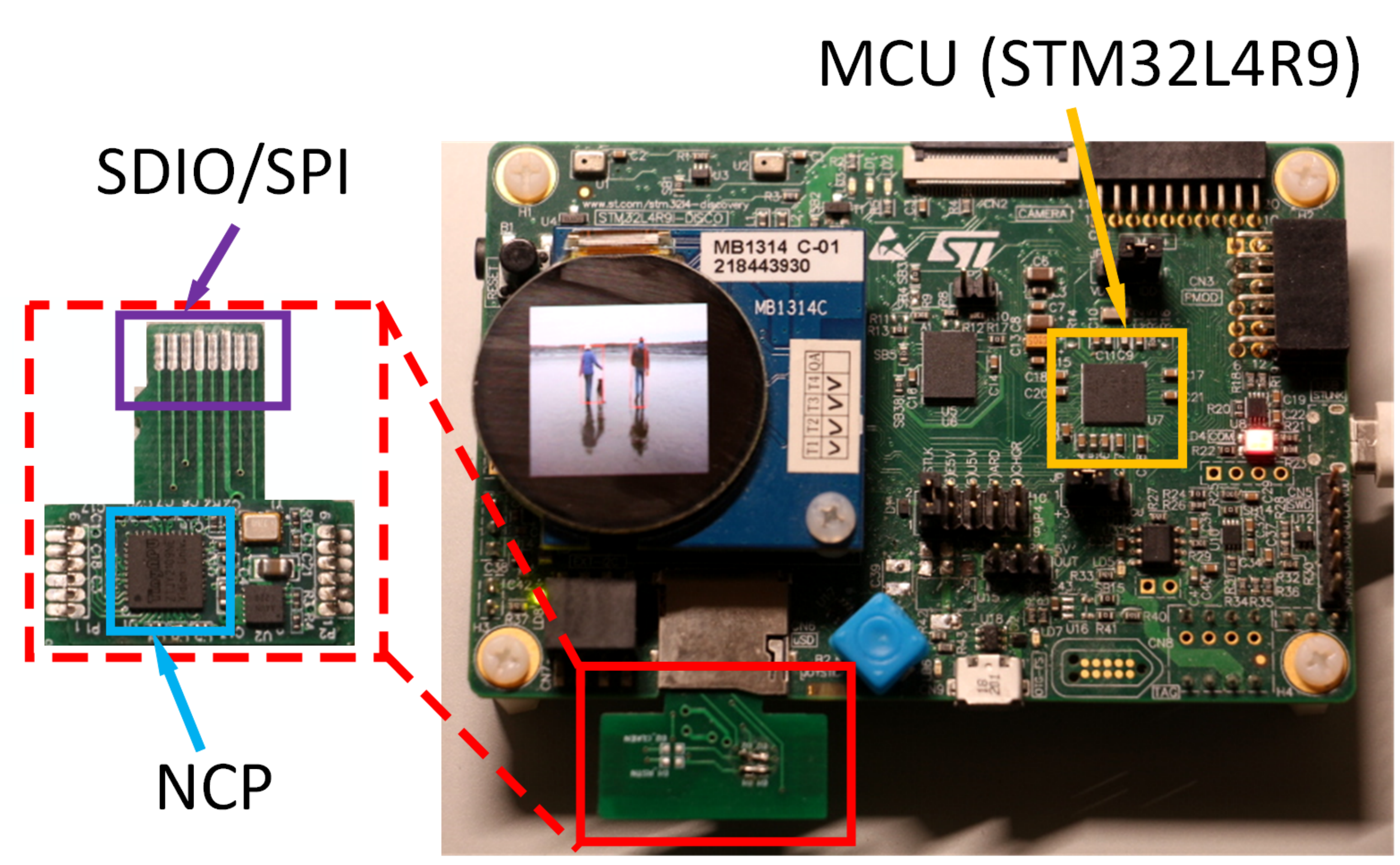}
	\end{center}
	\vspace{-0.5cm}
	\caption{The TinyML system for verification.}
	\vspace{-0.3cm}
	\label{fig:board}
\end{figure}

Fig.\ref{fig:board} shows the prototype verification system only consisting of STM32L4R9 MCU and our proposed NCP. Thanks to the innovative model (EtinyNet), co-processor (NCP) and application specific instruction-set, the entire system yields both of efficiency and flexibilty.
\section{Parameter-efficient EtinyNet Model} Since NCP handles CNN workloads entirely on-chip for pursuing extreme efficiency, we focus on reducing the model size for satisfying the memory constrains of IoT devices in TinyML, which is totally different from MobileML targeting at the reduction of MAdds. By presenting Linear Depthwise Block (LB) and Dense Linear Depthwise Block (DLB), we derive an extremely tiny CNN backbone EtinyNet, shown in Fig.\ref{fig:model}.
\subsection{Design of Proposed Blocks}

We present the linear depthwise convolution by removing the ReLU behind \emph{DWConv} of $\phi_{d1}$ under the observation that this non-linearity harms accuracy in the design of extremely parameter-efficient architectures, forming a specific case of sparse coding. Then, we introduce additional \emph{DWConv} of $\phi_{d2}$ behind \emph{PWConv} of $\phi_{p}$ to build a novel linear depthwise block (LB) by utilizing \emph{DWConv}'s parameter efficiency~\cite{etinynet}. The LB is defined as
\begin{equation}
	\mathbf{O} = \sigma(\phi_{d2}(\sigma(\phi_p(\phi_{d1}(\mathbf{I})))))\\
	\label{eq:conv}
\end{equation}

As shown in Fig~\ref{fig:model}(a), the structure of proposed LB can be represented as \emph{DWConv-PWConv-DWConv}, which is apparently different from the commonly used bottleneck block of \emph{PWConv-DWConv-PWConv} in mobile models, explained by the fact that increasing the proportion of \emph{DWConv} is beneficial to the accuracy of tiny models.

Additionally, we introduce the dense connection into LB for increasing its equivalent width, which is important and necessary for a higher accuracy~\cite{wideresnet}, as well as the very limited size of features and weights. We refer the resulting block to Dense Linear Depthwise Block (DLB) depicted in Fig~\ref{fig:model}(b). Note that we take the $\phi_{d1}$ and $\phi_{p}$ as a whole due to the removal of ReLU, and add the shortcut connection at the ends of these two layers.

\subsection{Architecture of EtinyNet Backbone}
\begin{figure}[t]
	\begin{center}
		\includegraphics[width=1 \linewidth]{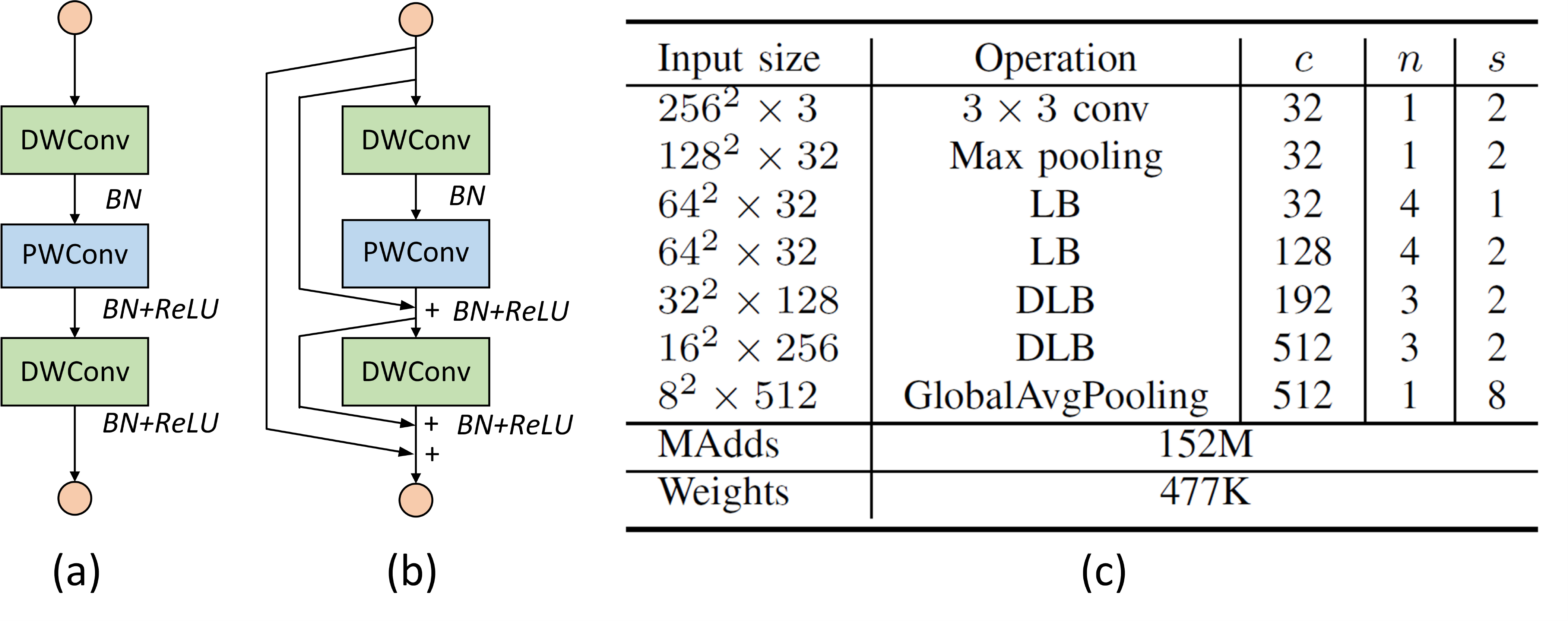}
	\end{center}
	\caption{The proposed building blocks that make up the EtinyNet. (a) is the Linear Depthwise Block (LB), (b) is the Dense Linear Depthwise Block (DLB), and (c) is the configuration of the backbone.}
	\label{fig:model}
\end{figure}

By stacking LBs and DLBs, we configure the EtinyNet backbone as indicated in Fig~\ref{fig:model}(c), where $n$, $c$ and $s$ represent block repeated times, the number of output channels, and the first layer's stride in each block (other layers' stride equaling one) respectively. Since dense connection consumes more memory space, we only utilize DLB at high level stages with much smaller feature maps. It's encouraging that EtinyNet backbone has only 477KB parameters and still achieves 66.5\% ImageNet Top-1 accuracy. The extreme compactness of EtinyNet makes it possible to design small footprint NCP that could run without off-chip DRAM.

\section{Application Specific Instruction-set for NCP} For easily deploying tiny CNN models on NCP, we define an application specific instruction-set. As shown in Table~\ref{tab:insset}, the set contains 13 instructions, belonging to neural operation type and control type respectively. It includes basic operations for tiny CNN models, and each instruction consists of 128 bits: 5 bits for operation code, and the rest for attributes of operations and operands. With each neural type instruction encoding an entire layer, the proposed instruction-set has a relatively coarser granularity, which simplifies the control complexity of hardware. Moreover, the basic operations included in the instruction-set provide sufficient ability to execute commonly-used CNN architectures (e.g., MobileNetV2~\cite{mobilenetv2}, MobileNeXt~\cite{mobilenext}, ~\emph{etc}). 
\begin{table}[h]\scriptsize
\caption{INSTRUCTION SET FOR PROPOSED NCP}
\label{tab:insset}
\centering
\begin{tabular}{|l|l|c|}
\hline
Instruction format              &    
  Description                           
      & Type   \\
\hline
\texttt{\textbf{bn} }           & \texttt{batch normalization}                    & N \\
\hline
\texttt{\textbf{relu}}          & \texttt{non-linear activation operation}        & N \\
\hline
\texttt{\textbf{conv}}          & \texttt{1x1 and 3x3 convolution \& bn, relu}   & N  \\
\hline
\texttt{\textbf{dwconv}}        & \texttt{3x3 depthwise conv \& bn, relu} & N \\
\hline
\texttt{\textbf{add}}           & \texttt{elementwise addition}                   & N \\
\hline
\texttt{\textbf{move} }         & \texttt{move tenor to target address}           & N \\
\hline
\texttt{\textbf{dsam}}          & \texttt{down-sampleing by factor of 2}          & N \\
\hline
\texttt{\textbf{usam}}          & \texttt{up-sampleing by factor of 2 }           & N \\
\hline
\texttt{\textbf{maxp}}          & \texttt{max pooling by factor of 2}             & N \\
\hline
\texttt{\textbf{gap}}           & \texttt{global average pooling}                 & N \\
\hline
\texttt{\textbf{jump}}          & \texttt{set program counter (PC) to target}     & C \\
\hline
\texttt{\textbf{sup}}           & \texttt{suspend processer}                      & C \\
\hline
\texttt{\textbf{end}}           & \texttt{suspend processer and reset PC}         & C \\
\hline
\end{tabular}
\end{table}

\section{Design of Neural Co-processor} 
\begin{figure}[h]
	\begin{center}
		\includegraphics[width=1 \linewidth]{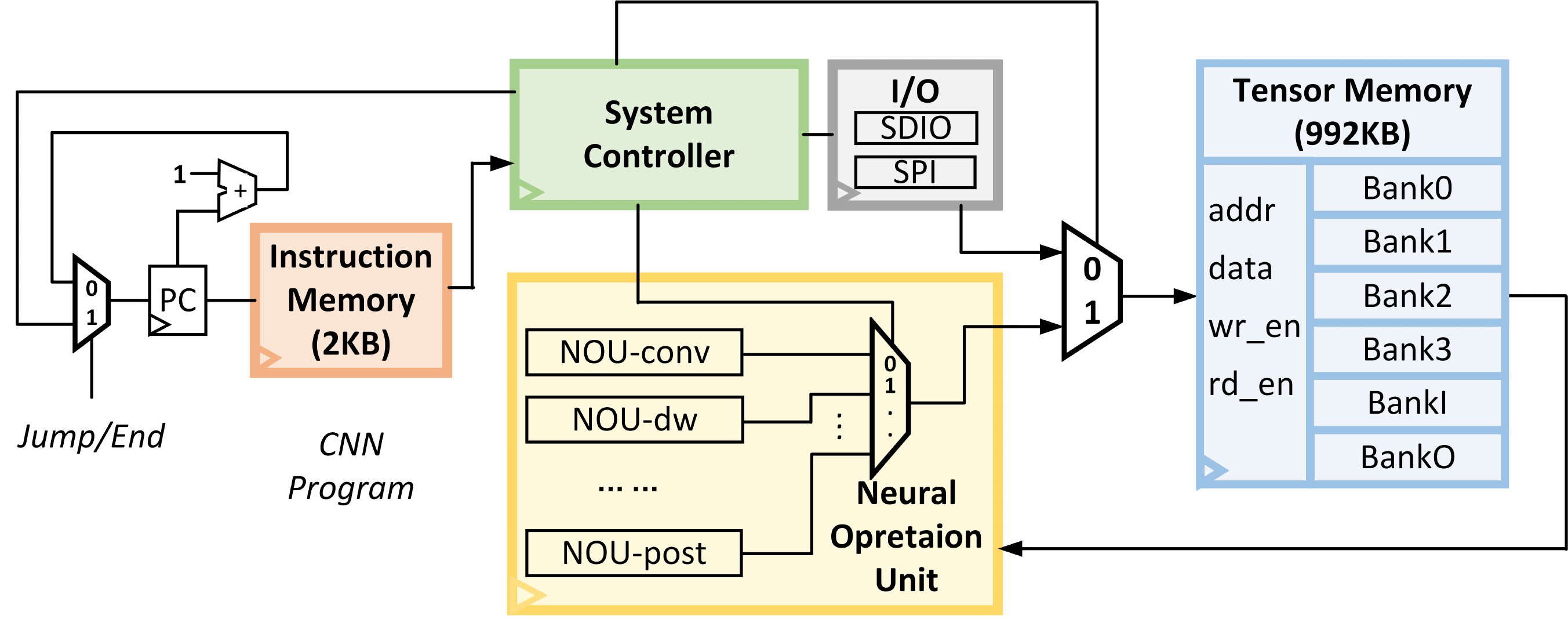}
	\end{center}
	\caption{The overall block diagram of the proposed NCP.}
	\label{fig:overview}
\end{figure}

As shown in Fig.\ref{fig:overview}, the proposed NCP consists of five main components: Neural Operation Unit (NOU), Tensor Memory (TM), Instruction Memory (IM), I/O and System Controller (SC). When NCP works, SC decodes one instruction fetched from IM and informs the NOU to start computing with decoded signals. The computing process takes multiple cycles, during which NOU reads operands from TM and writes results back automatically. Once completing the writing back process, SC continues to process the next instruction until an \textbf{\texttt{end}} or \textbf{\texttt{suspend}} instruction is encountered. When NOU is idle, TM is accessed through I/O. We will fully describe each component in the following parts.
\subsection{Neural Operation Unit} CNN workloads mainly come from operations of int8 \texttt{\textbf{conv}}, \texttt{\textbf{dwconv}} and float32 \texttt{\textbf{bn}}. To achieve a high energy efficiency, we respectively design special hardware units, termed NOU-conv, NOU-dw and NOU-post, focusing on optimizing the implementation of each operation. Furthermore, we deal with the design details in the following three aspects.

1) Different from other designs~\cite{convaix,vega} with fine grained instructions, we implement NOU-conv with a hardwired matrix multiply-accumulate (MAC)~\cite{mma} array, which helps to improve efficiency with a simpler control logic. The MAC array is designed to perform matrix outer product with parallelism in spatial and output channel dimension for handling the most computational costly \emph{3$\times$3 Conv} and \emph{PWConv} with \emph{im2col} operation. In this way, the number of effective multiplications in each cycle is fixed to $T_{oc} \times T_{hw}$. Note that the number of channels varies across different convolution layers, which may lead to inefficient computation for other ways of implementation (e.g., dot product). Conversely, our implementation manner can avoid the above-mentioned problem and improve the overall efficiency in running \emph{PWConv} of entire network. Moreover, the addition is realized by simple accumulation process instead of commonly-used adder tree with extra hardware overhead.

2) As for the implementation of \emph{DWConv}, the above designed MAC array proves its efficiency only in diagonal units. Given this, we turn to the classical convolution processing pipeline~\cite{diannao}, where nine multipliers and eight adders are arranged to compute \emph{DWConv} in each channel.  
The independence between channels allows us to extend pipelines easily, implementing a parallelism of $T_{oc}$ to build NOU-dw. Since the feature length in spatial dimension is usually much larger than the pipeline depth, the \emph{DWConv} can be performed in a fully pipelined manner, which yields NOU-dw an ultra high efficiency up to nearly 100\% .

3) In NOU-post unit, modules of int2float, float32 multiply-add, float2int and ReLU are designed and interconnected to perform post-operations of float32 BN, ReLU and element-wise addition. To reduce memory access as much as possible, multiplexers are further utilized to select data from the output of NOU-conv, NOU-dw or TM, and connect modules as needed, allowing flexible fusion of post-operations with the previous convolution layer. By implementing $T_{oc}$ pipelines to match the throughput of convolution, we effectively maximize the efficiency of fusion operations. 

\subsection{Tensor Memory and Tensor Layout} 1) TM is a single-port SRAM consisting of 6 banks, whose width is $ T_{tm} \times 8 $ bits, as shown in Fig~\ref{fig:overview}. Thanks to the compactness of EtinyNet, NCP only requires totally 992KB on-chip SRAM. The BankI (192KB) is responsible for caching $256\times256$ input RGB images. The 128KB sized Bank0 and Bank1 are arranged for caching feature maps, while Bank2 and Bank3 with a larger size of 256KB are used for storing weights. The BankO (32KB) is used to store final results, such as feature vectors and bonding boxes, \emph{etc}. TM's small capacity and simple structure yield our NCP a small footprint.

\begin{figure}[h]
	\begin{center}
		\includegraphics[width=1.0 \linewidth]{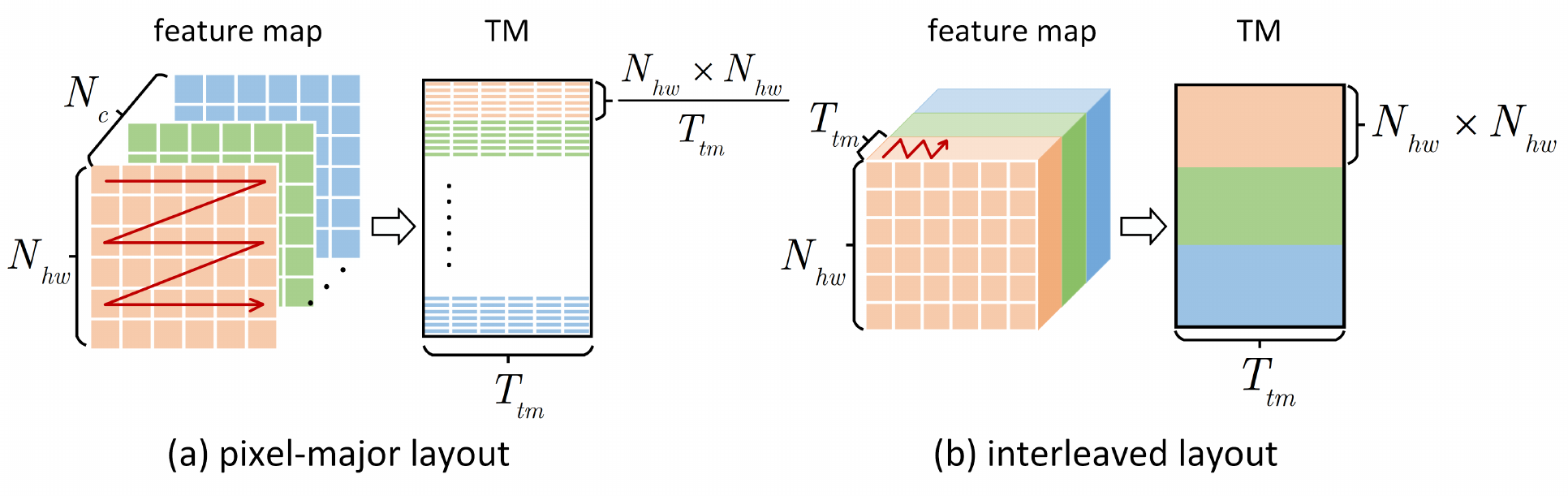}
	\end{center}
	\caption{Illustration of different tensor layouts. (a) Pixel-major layout. (b) Interleaved layout.}
	\label{fig:layout}
\end{figure}

2) The highly efficient NOU brings 2 types of tensor layouts, named pixel-major layout and interleaved layout respectively shown in Fig.\ref{fig:layout}. For the former, all pixels of the first channel are sequentially mapped to TM in a row-major order. Then, the next channel's counterpart is arranged in the same pattern until the last channel's pixels of a tensor are stored. For the latter, the whole tensor is divided into $N_{c} // T_{tm}$ tiles and are placed in TM sequentially, while each tile is arranged in a channel-major order. Different layouts are required for NOUs to achieve the maximum efficiency. For example, the input of NOU-conv prefers pixel-major layout because spatially continuous $T_{hw}$ pixels of a channel need to be multiplied and added at a time by MAC array, while the reverse is the case for NOU-dw.

\begin{figure}[h]
	\begin{center}
		\includegraphics[width=1.0 \linewidth]{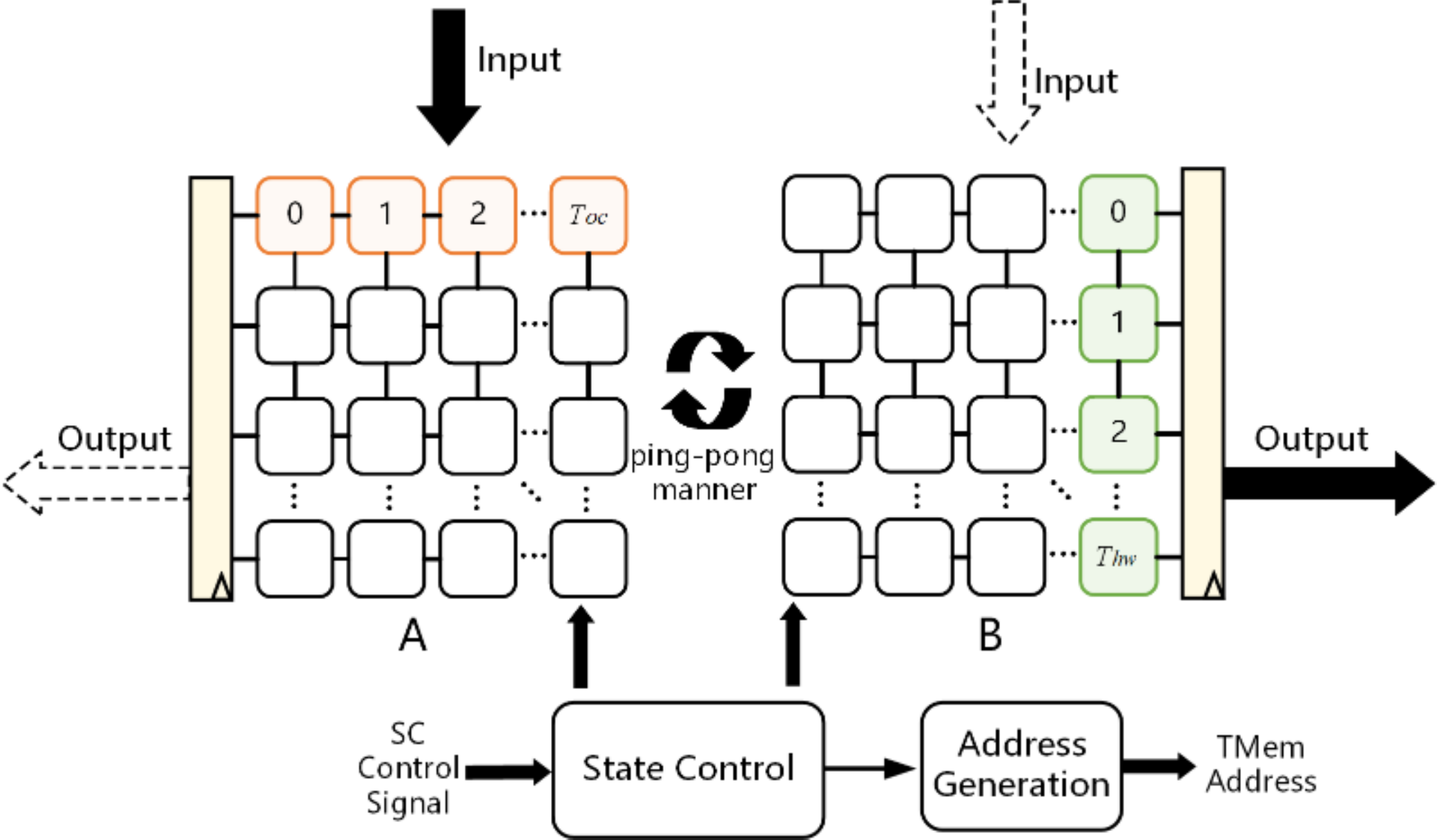}
	\end{center}
	\caption{The proposed efficient layout conversion circuit.}
	\label{fig:lcc}
\end{figure}

3) Running the proposed LB, DLB, and other blocks with NOU, the layout between adjacent \emph{DWConv} and \emph{PWConv} is constantly varying, which seriously decreases the computing efficiency because of the discontinuous memory access. It takes NOU-conv $T_{oc}$ times to read the output of NOU-dw stored in an interleaved layout for performing a single matrix outer product operation. Hence, an efficient layout conversion circuit is designed to tackle this problem. As shown in Fig.\ref{fig:lcc}, the circuit is composed of two $T_{oc} \times T_{hw}$ register arrays A and B, working in a ping-pong mechanism. At the beginning, array A receives $T_{oc}$ inputs at a time, after $T_{hw}$ cycles, A will be filled and start to output $T_{hw}$ results at a time in the transposed dimension. Since reading A empty requires $T_{oc}$ cycles, the new coming data to be converted will be sent to array B in order to maintain the pipeline. When B is full and A completes the readout, the role of them are exchanged. This strategy obviously boosts the efficiency of valid memory access for computing.

\subsection{Characteristics}We implement our NCP using TSMC 65nm low-power technology. While $T_{tm}=32$, $T_{oc}=16$ and $T_{hw}=32$, NCP contains 512 of 8-bit MACs in NOU-conv, 144 of 8-bit multipliers and 16 of adder trees in NOU-dw, and 16 of float32 MACs in NOU-post. When running at the maximum frequency of 250MHz, NOU-conv and NOU-post are active every cycle, achieving a peak performance of 264 GOP/s. 
\section{Experimental Results} 
\subsection{EtinyNet Evaluation}Table~\ref{tab:models} lists the ImageNet-1000 classification results of well-known lightweight CNNs, including MobileNetV2~\cite{mobilenetv2}, MobileNeXt~\cite{mobilenext}, ShuffleNetV2~\cite{shufflenetv2}, and MCUNet series~\cite{mcunetv2}. We pay more attention to the backbone because the fully-connected layer is generally not involved in most of visual models. Among these competitive models, MCUNet gets the highest accuracy at the cost of model size up to 2048K. Compared with tiny models in similar size, our EtinyNet reaches 66.5\% top-1 and 86.8\% top-5 accuracy, outperforming the most competitive MCUNetV2-M4 by significant 1.6\% top-1 accuracy. Moreover, EtinyNet-0.75, the width of each layer is shrunk by 0.75, outperforms MCUNet-320kB by significant 2.6\% top-1 accuracy with 60K fewer parameters. Obviously, EtinyNet yields much higher accuracy at the same level of storage consumption, and is more suitable for TinyML systems.
\begin{table}[h]
\caption{Comparison of state-of-the-art tiny models over accuracy on ImageNet. "B" denotes backbone. "-" denotes not reported. }
\label{tab:models}
\centering
\begin{tabular}{l||c|c|c}
\toprule
Model                  & \#Params. (K)      & Top-1 Acc.      & Top-5 Acc.           \\
\midrule
MobileNeXt-0.35        & 812(B) / 1836     & 64.7            & 85.7                 \\
MobileNetV2-0.35       & 740(B) / 1764     & 60.3            & 82.9                 \\
ShuffleNetV2-0.5       & 566(B) / 1590     & 61.1            & 82.6                 \\
\hline
MCUNet                 & -(B)   / 2048 & \textbf{70.7}            & -                    \\
MCUNet-320kB           & -(B)   / 740      & 61.8            & 84.2                 \\
MCUNetV2-M4            & -(B)   / 1034 & 64.9            & 86.2                 \\
\hline
EtinyNet               & 477(B) / \textbf{989}      & \textbf{66.5}            & \textbf{86.8}                 \\
EtinyNet-0.75          & 296(B) / \textbf{680}      & \textbf{64.4}            & \textbf{85.2}                 \\
EtinyNet-0.5           & 126(B) / 446      & 59.3            & 81.2                 \\
\bottomrule
\end{tabular}
\end{table}
\subsection{NCP Evaluation} 
\begin{table}[h]\scriptsize
	\caption{Comparison with state-of-the-art neural processors. "-" denotes not reported.}
	\label{tab:asip}
	\centering
	\begin{tabular}{l||c|c|c|c|c}
		\toprule
		Component                  & NullHop                   & ConvAix               & YodaNN            & Vega & NCP     \\
		\hline
		Technology             & 28nm                        & 28nm                    & 65nm                 & 22nm  & 65nm     \\
		Area ($mm^2$)         & 6.3                      & 3.53                  & 1.9                & 12    & 10.88    \\
		DRAM Used                  & yes                      & yes                   & none               & none  & none     \\
		FC Support                 & none                     & yes                   & none               & yes & none     \\
		\hline
		CNN model   & VGG16    & MbV1  &VGG19    & RVGGA0 & EtinyNet \\
		ImageNet Acc.          & 68.3\%                   & 70.6\%                  & -                   & 72.4\%  & 66.5\%    \\
		Latency                & 72.9ms                     & 14.2ms                  & 75.2ms             & 118ms   & \textbf{5.5ms}        \\
		\hline
		Typ. Power            & \multirow{2}{*}{155.0}  & \multirow{2}{*}{313.1}    & \multirow{2}{*}{153}   & \multirow{2}{*}{37.3}  & \multirow{2}{*}{73.6}     \\
		(mW)                	 &                          &                        &                       &                   &                          \\ 
		\hline
		Peak Perf.          & \multirow{2}{*}{128}   & \multirow{2}{*}{262.6} & \multirow{2}{*}{1500} & \multirow{2}{*}{32.2} & \multirow{2}{*}{264}   \\ 
		(GOP/s)                	 &                          &                        &                       &                   &                          \\ 
		\hline
		Energy Eff.         & \multirow{2}{*}{2714.8}  & \multirow{2}{*}{256.3} & \multirow{2}{*}{8500} & \multirow{2}{*}{631.4} & \multirow{2}{*}{751.0}    \\
		(GOP/s/W)         &                          &                        &                       &                   &                          \\
		\hline
		Processing Eff.      & \multirow{2}{*}{1.21}    & \multirow{2}{*}{15.18} & \multirow{2}{*}{2.95}  &\multirow{2}{*}{1.93} & \multirow{2}{*}{\textbf{449.1}}    \\
		(Frames/s/mJ)           &                          &                        &                       &                     &                           \\
		\bottomrule
	\end{tabular}
\end{table}

As shown in Table~\ref{tab:asip}, running general CNN models usually needs DRAMs to store their enormous weights and features~\cite{convaix,nullhop}, resulting in considerable power consumption and processing latency. As for no DRAM access methods, YodaNN~\cite{yodann} yields the highest peak performance and energy efficiency, but it is a dedicated accelerator only for binarized networks with very limited accuracy. Except that, Vega~\cite{vega} gets the lowest power and the maximum latency, which leads to the lowest peak performance. To comprehensively assess the throughput, energy consumption and speed of various neural processors in TinyML application, we prefer to use the metric of processing efficiency, which is the number of frames processed per unit time and per unit power consumption. Our proposed NCP reaches an extremely high processing efficiency up to 449.1 Frames/s/mJ, at least $29 \times$ higher than other solutions, suggesting the unique superiority of NCP in this particular field. As for the reason, the specially designed NOU, tenser layout and coarse-grained instruction-set jointly decrease the delay and the power of inference.

\subsection{TinyML System Verification}
We compare our proposed system with existing prominent MCU-based TinyML systems. As shown in Table~\ref{tab:mcu}, CMSIS-NN obtains 59.5\% ImageNet accuracy at 2FPS, promoted by MCUNet to 5FPS at the expense of accuracy dropping to 49.9\%. In comparison, our solution reaches up to 66.5\% accuracy and 30FPS, achieving the goal of real-time visual processing in TinyML. Furthermore, since existing methods burden MCUs with entire CNN models, high-performance MCUs (STM32H743/STM32F746) running at the upper-limit frequency (480MHz/216MHz) are necessary. 
Although flexible, general-purpose MCU is of low energy efficiency in computing massive tensors, which results in considerable power consumption up to about 600mW. In contrast, the proposed solution allows us to perform the same flexible task only with a low-end MCU (STM32L4R9,120MHz) and proposed NCP, which boosts the energy efficiency of the entire system and achieves an ultra-low power of 160mW.
\begin{table}[H]
	\caption{Comparison with MCU-based designs on image classification (cls) and object detection (det). $\ast$ denotes reproduced results.}
	\label{tab:mcu}
	\centering
	\begin{tabular}{l|l|c|c|c|c}
		\toprule
		& Method     & Hardware   & Acc/mAP          & FPS          & Power           \\
		\hline
		\multirow{3}{*}{Cls}         & CMSIS-NN    & H743       & 59.5\%           & 2            & $\ast$675 mW           \\
		& MCUNet      & F746       & 49.9\%           & 5            & $\ast$525 mW    \\
		& Ours        & L4R9+NCP   & \textbf{66.5\%}  & \textbf{30}  & \textbf{160 mW} \\
		\hline
		\multirow{3}{*}{Det}         & CMSIS-NN    & H743       & 31.6\%           & 10           & $\ast$640 mW    \\
		& MCUNet      & H743       & 51.4\%           & 3            & $\ast$650 mW    \\
		& Ours        & L4R9+NCP   & \textbf{56.4\%}  & \textbf{30}  & \textbf{160 mW} \\
		\bottomrule
	\end{tabular}
\end{table}

In addition, we benchmark the object detection performance on Pascal VOC dataset. The results indicate that our system also greatly improves its performance, which makes AIoT more promising in extensive applications.
\section{Conclusion}
In this brief, we propose an ultra-low power TinyML system for real-time visual processing by designing 1) an extremely tiny CNN backbone EtinyNet, 2) an ASIC-based neural co-processor and 3) an application specific instruction-set. Our study greatly advances the TinyML community and promises to drastically expand the application scope of AIoT.

\bibliographystyle{IEEEtran}


\end{document}